\documentclass[preprintnumbers,amsmath,amssymbm,prd]{revtex4}
\usepackage{epsfig}
\usepackage{graphicx}

\begin{document}
\title{The entropy emission properties of near-extremal Reissner-Nordstr\"om black holes}
\author{Shahar Hod}
\address{The Ruppin Academic Center, Emeq Hefer 40250, Israel}
\address{ }
\address{The Hadassah Institute, Jerusalem 91010, Israel}
\date{\today}

\begin{abstract}
\ \ \ Bekenstein and Mayo have revealed an interesting property of
evaporating $(3+1)$-dimensional Schwarzschild black holes: their
entropy emission rates $\dot S_{\text{Sch}}$ are related to their
energy emission rates $P$ by the simple relation $\dot
S_{\text{Sch}}=C_{\text{Sch}}\times (P/\hbar)^{1/2}$, where
$C_{\text{Sch}}$ is a numerically computed dimensionless
coefficient. Remembering that $(1+1)$-dimensional perfect black-body
emitters are characterized by the same functional relation, $\dot
S^{1+1}=C^{1+1}\times(P/\hbar)^{1/2}$ [with
$C^{1+1}=(\pi/3)^{1/2}$], Bekenstein and Mayo have concluded that,
in their entropy emission properties, $(3+1)$-dimensional
Schwarzschild black holes behave effectively as $(1+1)$-dimensional
entropy emitters. Later studies have shown that this intriguing
property is actually a generic feature of all radiating
$(D+1)$-dimensional Schwarzschild black holes. One naturally wonders
whether all black holes behave as simple $(1+1)$-dimensional entropy
emitters? In order to address this interesting question, we shall
study in this paper the entropy emission properties of
Reissner-Nordstr\"om black holes. We shall show, in particular, that
the physical properties which characterize the neutral sector of the
Hawking emission spectra of these black holes can be studied {\it
analytically} in the near-extremal $T_{\text{BH}}\to0$ regime (here
$T_{\text{BH}}$ is the Bekenstein-Hawking temperature of the black
hole). We find that the Hawking radiation spectra of massless
neutral scalar fields and coupled electromagnetic-gravitational
fields are characterized by the non-trivial entropy-energy relations
$\dot
S^{\text{Scalar}}_{\text{RN}}=-C^{\text{Scalar}}_{\text{RN}}\times
(AP^3/\hbar^3)^{1/4}\ln(AP/\hbar)$ and $\dot
S^{\text{Elec-Grav}}_{\text{RN}}=-C^{\text{Elec-Grav}}_{\text{RN}}\times
(A^4P^9/\hbar^9)^{1/10}\ln(AP/\hbar)$ in the near-extremal
$T_{\text{BH}}\to0$ limit [here
$\{C^{\text{Scalar}}_{\text{RN}},C^{\text{Elec-Grav}}_{\text{RN}}\}$
are analytically calculated dimensionless coefficients and $A$ is
the surface area of the Reissner-Nordstr\"om black hole]. Our
analytical results therefore indicate that {\it not} all black holes
behave as simple $(1+1)$-dimensional entropy emitters.
\end{abstract}
\bigskip
\maketitle


\section{Introduction}

The entropy emission rate $\dot S$ and the energy emission rate
(power) $P$ of a perfect black-body (BB) emitter in a flat
$(3+1)$-dimensional spacetime are related by the well known
Stefan-Boltzmann radiation law \cite{Allen} (we use gravitational
units in which $G=c=k_{\text{B}}=1$)
\begin{equation}\label{Eq1}
\dot
S^{3+1}_{\text{BB}}=C^{3+1}\times\Big({{AP^3}\over{\hbar^3}}\Big)^{1/4}\
,
\end{equation}
where $A$ is the surface area of the $(3+1)$-dimensional radiating
black-body and $C^{3+1}=({{32\pi^2}/{1215}})^{1/4}$ is a
dimensionless proportionality coefficient.

However, in a very interesting work, Bekenstein and Mayo
\cite{BekMay} have revealed the remarkable fact that the Hawking
radiation spectra \cite{Haw1} of $(3+1)$-dimensional Schwarzschild
black holes are characterized by the qualitatively different (and
mathematically much simpler) entropy-energy relation \cite{BekMay}
\begin{equation}\label{Eq2} \dot
S^{3+1}_{\text{Sch}}=C^{3+1}_{\text{Sch}}\times\Big({{P}\over{\hbar}}\Big)^{1/2}\
,
\end{equation}
where $C^{3+1}_{\text{Sch}}$ is a numerically computed coefficient
\cite{BekMay} which depends on the characteristic greybody factors
\cite{Page} of the $(3+1)$-dimensional Schwarzschild black-hole
spacetime.

Bekenstein and Mayo \cite{BekMay} have emphasized the interesting
fact that the entropy-energy relation (\ref{Eq2}), which
characterizes the Hawking emission spectra of $(3+1)$-dimensional
Schwarzschild black holes, has the same functional form as the
entropy-energy relation \cite{BekMay}
\begin{equation}\label{Eq3}
\dot
S^{1+1}_{\text{BB}}=C^{1+1}_{\text{BB}}\times\Big({{P}\over{\hbar}}\Big)^{1/2}
\end{equation}
which characterizes the emission spectra of $(1+1)$-dimensional
perfect black-body emitters [one finds
$C^{1+1}_{\text{BB}}=(\pi/3)^{1/2}$ for a $(1+1)$-dimensional
perfect black-body emitter  \cite{BekMay}]. Hence, Bekenstein and
Mayo \cite{BekMay} have reached the intriguing conclusion that, in
their entropy emission properties, $(3+1)$-dimensional Schwarzschild
black holes behave effectively as $(1+1)$-dimensional [and not as
$(3+1)$-dimensional] thermal entropy emitters [see Eqs. (\ref{Eq2})
and (\ref{Eq3})]. It is worth noting that it was later proved
\cite{Hodj,Mirj} that this intriguing property of the
$(3+1)$-dimensional Schwarzschild black holes is actually a generic
characteristic of all radiating $(D+1)$-dimensional Schwarzschild
black holes.

One naturally wonders whether this intriguing physical property of
the Schwarzschild black holes is shared by all black holes? In
particular, we raise here the following question: do all radiating
black holes behave as simple $(1+1)$-dimensional entropy emitters?
In order to address this interesting question, we shall analyze in
this paper the entropy emission properties of Reissner-Nordstr\"om
black holes. As we shall show below, the physical properties which
characterize the neutral sector of the Hawking radiation spectra of
these black holes can be studied {\it analytically} in the
near-extremal $T_{\text{BH}}\to0$ regime [here $T_{\text{BH}}$ is
the Bekenstein-Hawking temperature of the Reissner-Nordstr\"om black
hole, see Eq. (\ref{Eq6}) below]. Our analytical results (to be
presented below) indicate that {\it not} all radiating black holes
behave as simple $(1+1)$-dimensional entropy emitters.

\section{The Hawking radiation spectra of near-extremal Reissner-Nordstr\"om black
holes}

In the present section we shall study the Hawking emission of
massless neutral fields from near-extremal Reissner-Nordstr\"om
black holes. The semi-classical Hawking radiation power
$P_{\text{RN}}$ and the semi-classical entropy emission rate $\dot
S_{\text{RN}}$ for one bosonic degree of freedom are given
respectively by the integral relations \cite{Page,Zurek,Bekf}
\begin{equation}\label{Eq4}
P_{\text{RN}}={{\hbar}\over{2\pi}}\sum_{l,m}\int_0^{\infty} {{\Gamma
\omega}\over{e^{\hbar\omega/T_{\text{BH}}}-1}}d\omega\  ,
\end{equation}
and
\begin{equation}\label{Eq5}
\dot
S_{\text{RN}}={{1}\over{2\pi}}\sum_{l,m}\int_0^{\infty}\Big[{{\Gamma}\over{e^{\hbar\omega/T_{\text{BH}}}-1}}
\ln\Big({{e^{\hbar\omega/T_{\text{BH}}}-1}\over{\Gamma}}+1\Big)+\ln\Big(1+{{\Gamma}
\over{e^{\hbar\omega/T_{\text{BH}}}-1}}\Big)\Big]d\omega\  ,
\end{equation}
where $\{l,m\}$ are the angular harmonic indices of the emitted
field mode, $\Gamma=\Gamma_{lm}(\omega)$ are the black-hole-field
greybody factors \cite{Page}, and
\begin{equation}\label{Eq6}
T_{\text{BH}}={{\hbar(r_+-r_-)}\over{4\pi r^2_+}}\
\end{equation}
with $r_{\pm}=M+(M^2-Q^2)^{1/2}$ is the semi-classical
Bekenstein-Hawking temperature of the Reissner-Nordstr\"om black
hole [here $\{r_+,r_-\}$ are the horizon radii of the
Reissner-Nordstr\"om black hole, and $\{M,Q\}$ are the black-hole
mass and charge, respectively].

The characteristic thermal factor
$\omega/(e^{\hbar\omega/T_{\text{BH}}}-1)$ that appears in the
expression (\ref{Eq4}) for the semi-classical black-hole radiation
power implies that the Hawking emission spectra peak at the
characteristic frequency
\begin{equation}\label{Eq7}
{{\hbar\omega^{\text{peak}}}\over{T_{\text{BH}}}}=O(1)\ .
\end{equation}
Furthermore, taking cognizance of the fact that the
Bekenstein-Hawking temperature (\ref{Eq6}) of a {\it near-extremal}
\cite{Noteex} Reissner-Nordstr\"om black hole is characterized by
the strong inequality
\begin{equation}\label{Eq8}
{{MT_{\text{BH}}}\over{\hbar}}\ll1\  ,
\end{equation}
one finds the closely related strong inequality
\begin{equation}\label{Eq9}
M\omega^{\text{peak}}\ll1\
\end{equation}
for the characteristic emission frequencies that constitute the
Hawking black-hole radiation spectra in the near-extremal regime
(\ref{Eq8}).

It is well known \cite{Page,Hodep} that the dimensionless greybody
factors $\Gamma_{lm}(\omega)$, which quantify the interaction of the
emitted field modes with the effective curvature barrier in the
exterior region of the black-hole spacetime, can be calculated
analytically in the low-frequency regime (\ref{Eq9}) [it is worth
emphasizing again that the low-frequency regime (\ref{Eq9})
dominates the neutral sector of the Hawking radiation spectra in the
low-temperature (near-extremal) regime (\ref{Eq8})]. We shall now
use this fact in order to study {\it analytically} the physical
properties which characterize the neutral sector of the Hawking
black-hole radiation spectra in the near-extremal (low-temperature)
limit (\ref{Eq8}).

\subsection{The Hawking emission of massless scalar quanta}

Following the analysis presented in \cite{Page}, one finds the
leading-order behavior
\begin{equation}\label{Eq10}
\Gamma_{lm}=\Big[{{(l!)^2}\over{(2l)!(2l+1)!!}}\Big]^2\prod_{n=1}^{l}\Big[1+\Big({{\hbar\omega}\over{2\pi
T_{\text{BH}}\cdot
n}}\Big)^2\Big]\Big({{AT_{\text{BH}}\omega}\over{\hbar}}\Big)^{2l}{{A\omega^2}\over{\pi}}
\cdot[1+O(AT_{\text{BH}}\omega/\hbar)^{2l+1}]\
\end{equation}
for the greybody factors which characterize the emission of scalar
quanta in the low-frequency regime (\ref{Eq9}). Here
\begin{equation}\label{Eq11}
A=4\pi r^2_+\
\end{equation}
is the surface area of the Reissner-Nordstr\"om black hole. From
(\ref{Eq10}) one finds that, in the low-frequency regime
(\ref{Eq9}), the scalar Hawking black-hole radiation spectrum is
dominated by the fundamental $l=m=0$ mode [it is worth emphasizing
the fact that the Hawking emission of scalar modes with $l>0$ is
suppressed as compared to the Hawking emission of the fundamental
$l=m=0$ scalar mode. This characteristic property of the Hawking
black-hole emission spectra stems from the fact that the greybody
factors of the higher scalar modes (that is, scalar modes which are
characterized by $l>0$) contain higher powers (as compared to the
fundamental $l=m=0$ scalar mode) of the small quantity $\omega
(r_+-r_-)\ll1$ [see Eq. (\ref{Eq10})]]. In particular, one finds
[see Eq. (\ref{Eq10})]
\begin{equation}\label{Eq12}
\Gamma_{00}={{A\omega^2}\over{\pi}}\cdot[1+O(AT_{\text{BH}}\omega/\hbar)]
\end{equation}
in the small frequency regime (\ref{Eq9}) which characterizes the
neutral sector of the Hawking emission spectra in the
low-temperature (near-extremal) regime (\ref{Eq8}).

Substituting (\ref{Eq12}) into Eqs. (\ref{Eq4}) and (\ref{Eq5}), one
finds after some algebra \cite{Notealg}
\begin{equation}\label{Eq13}
P_{\text{RN}}={{\pi^2}\over{30}}{{AT^4_{\text{BH}}}\over{\hbar^3}}
\end{equation}
and
\begin{equation}\label{Eq14}
\dot
S_{\text{RN}}=-{{\zeta(3)}\over{\pi^2}}{{AT^3_{\text{BH}}}\over{\hbar^{3}}}\cdot\Big[\ln\Big({{AT^2_{\text{BH}}}\over
{\hbar^2}}\Big)+O(1)\Big]
\end{equation}
for the scalar Hawking radiation power and the scalar Hawking
entropy emission rate of the near-extremal Reissner-Nordstr\"om
black holes.

Finally, substituting (\ref{Eq13}) into (\ref{Eq14}), one can
express the black-hole entropy emission rate in terms of the Hawking
radiation power:
\begin{equation}\label{Eq15}
\dot
S^{\text{Scalar}}_{\text{RN}}=-C^{\text{Scalar}}_{\text{RN}}\times
\Big({{AP^3_{\text{RN}}}\over{\hbar^3}}\Big)^{1/4}\ln\Big({{AP_{\text{RN}}}\over{\hbar}}\Big)\
,
\end{equation}
where the analytically calculated coefficient
$C^{\text{Scalar}}_{\text{RN}}$ is given by
\begin{equation}\label{Eq16}
C^{\text{Scalar}}_{\text{RN}}={{30^{3/4}\zeta(3)}\over{2\pi^{7/2}}}\
.
\end{equation}
It is worth emphasizing the fact that the entropy emission rate
(\ref{Eq15}), which characterizes the scalar Hawking emission
spectra of the near-extremal Reissner-Nordstr\"om black holes, does
{\it not} have the simple $(1+1)$-dimensional entropy-energy
functional relation (\ref{Eq2}) which characterizes the Hawking
emission spectra of the Schwarzschild black holes. It is important
to emphasize that Mirza, Oboudiat, and Zare \cite{Mirj} have reached
similar conclusions for 3-dimensional rotating BTZ black holes and
for D-dimensional Lovelock black holes in odd and even dimensions
(Refs. \cite{Hodj} and \cite{Mirj} have also considered the case of
D-dimensional general relativistic black holes).

\subsection{The Hawking emission of coupled electromagnetic-gravitational
quanta}

For the case of coupled electromagnetic-gravitational quanta, one
finds the leading-order behavior \cite{Hodep}
\begin{equation}\label{Eq17}
\Gamma_{11}=\Gamma_{2m}={4\over9}\Big({{A\omega^2}\over{4\pi}}\Big)^4
\end{equation}
for the greybody factors in the low-frequency regime (\ref{Eq9})
which characterizes the neutral sector of the Hawking black-hole
emission spectra in the near-extremal (low-temperature) regime
(\ref{Eq8}). [It is worth emphasizing the fact that the Hawking
emission of coupled electromagnetic-gravitational modes with $l>2$
is suppressed as compared to the Hawking emission of coupled
electromagnetic-gravitational modes with $l=1$ and $l=2$. This
characteristic property of the Hawking black-hole emission spectra
stems from the fact that the greybody factors of coupled
electromagnetic-gravitational modes with higher-$l$ values (that is,
electromagnetic-gravitational modes which are characterized by
$l>2$) contain higher powers (as compared to the $l=1$ and $l=2$
modes) of the small quantity $\omega r_+\ll1$ \cite{Hodep}].

Substituting (\ref{Eq17}) into Eqs. (\ref{Eq4}) and (\ref{Eq5}), one
finds after some algebra \cite{Notealg2}
\begin{equation}\label{Eq18}
P_{\text{RN}}={{4\pi^5}\over{297}}{{A^4T^{10}_{\text{BH}}}\over{\hbar^9}}
\end{equation}
and
\begin{equation}\label{Eq19}
\dot
S_{\text{RN}}=-{{560\zeta(9)}\over{\pi^5}}
{{A^4T^9_{\text{BH}}}\over{\hbar^{9}}}\cdot\Big[\ln\Big({{AT^2_{\text{BH}}}\over
{\hbar^2}}\Big)+O(1)\Big]
\end{equation}
for the electromagnetic-gravitational Hawking radiation power and
the electromagnetic-gravitational Hawking entropy emission rate of
the near-extremal Reissner-Nordstr\"om black holes.

Finally, substituting (\ref{Eq18}) into (\ref{Eq19}), one can
express the black-hole entropy emission rate in terms of the Hawking
radiation power:
\begin{equation}\label{Eq20}
\dot
S^{\text{Elec-Grav}}_{\text{RN}}=-C^{\text{Elec-Grav}}_{\text{RN}}\times
\Big({{A^4P^9_{\text{RN}}}\over{\hbar^9}}\Big)^{1/10}\ln\Big({{AP_{\text{RN}}}\over{\hbar}}\Big)\
,
\end{equation}
where the analytically calculated coefficient
$C^{\text{Elec-Grav}}_{\text{RN}}$ is given by
\begin{equation}\label{Eq21}
C^{\text{Elec-Grav}}_{\text{RN}}={{112\zeta(9)}\over{\pi^{19/2}}}\Big({{297}\over{4}}\Big)^{9/10}\
.
\end{equation}
It is worth emphasizing again that the entropy emission rate
(\ref{Eq20}), which characterizes the electromagnetic-gravitational
Hawking emission spectra of the near-extremal Reissner-Nordstr\"om
black holes, does {\it not} have the simple $(1+1)$-dimensional
entropy-energy functional relation (\ref{Eq2}) which characterizes
the Hawking radiation spectra of the Schwarzschild black holes.

\section{Summary}

In a very interesting paper \cite{BekMay}, Bekenstein and Mayo have
revealed that, in their entropy emission properties,
$(3+1)$-dimensional Schwarzschild black holes behave effectively as
$(1+1)$-dimensional [and not as $(3+1)$-dimensional] thermal entropy
emitters [see Eqs. (\ref{Eq2}) and (\ref{Eq3})]. Later studies
\cite{Hodj,Mirj} have extended the analysis of \cite{BekMay} to
higher dimensional black holes and proved that all radiating
$(D+1)$-dimensional Schwarzschild black holes are characterized by
this intriguing physical property.

One naturally wonders whether this interesting property of the
Schwarzschild black holes is shared by all black holes? In
particular, motivated by the results of \cite{BekMay,Hodj,Mirj}, we
have raised here the following question: do all radiating black
holes behave as simple $(1+1)$-dimensional entropy emitters? In
order to address this intriguing question, we have explored in this
paper the entropy emission properties of Reissner-Nordstr\"om black
holes. In particular, we have shown that the physical properties
which characterize the neutral sector of the Hawking emission
spectra of these black holes can be studied {\it analytically} in
the low-temperature (near-extremal) $T_{\text{BH}}\to0$ regime.

We have explicitly shown that the analytically derived expressions
for the Hawking entropy emission rates of massless scalar fields and
coupled electromagnetic-gravitational fields by near-extremal
Reissner-Nordstr\"om black holes [see Eqs. (\ref{Eq15}) and
(\ref{Eq20})] do not have the simple $(1+1)$-dimensional
entropy-energy functional relation (\ref{Eq2}) which characterizes
the Hawking emission spectra of the Schwarzschild black holes. Our
analytical results therefore indicate that {\it not} all black holes
behave as simple $(1+1)$-dimensional entropy emitters.

\bigskip
\noindent
{\bf ACKNOWLEDGMENTS}
\bigskip

This research is supported by the Carmel Science Foundation. I thank
Yael Oren, Arbel M. Ongo, Ayelet B. Lata, and Alona B. Tea for
stimulating discussions.



\begin{thebibliography}{99}
\bibitem{Noteex} Note that near-extremal black holes are
characterized by the property $(r_+-r_-)/r_+\ll1$, which implies
$MT_{\text{BH}}/\hbar\ll1$ [see Eq. (\ref{Eq6})].
\bibitem{Notealg} Here we have used the relations
${{e^{\hbar\omega/T_{\text{BH}}}-1}\over{A\omega^2/\pi}}\gg1$ and
$\ln\Big(1+{{A\omega^2/\pi}\over{e^{\hbar\omega/T_{\text{BH}}}-1}}\Big)\simeq
{{A\omega^2/\pi}\over{e^{\hbar\omega/T_{\text{BH}}}-1}}\ll1$ in the
low frequency regime $M\omega\ll1$ [see Eq. (\ref{Eq9})] which
characterizes the neutral sector of the Hawking black-hole emission
spectra in the near-extremal (low-temperature) regime (\ref{Eq8}).
\bibitem{Notealg2} Here we have used the relations
${{e^{\hbar\omega/T_{\text{BH}}}-1}\over{{4\over9}(A\omega^2/4\pi)^4}}\gg1$
and
$\ln\Big[1+{{{4\over9}(A\omega^2/4\pi)^4}\over{e^{\hbar\omega/T_{\text{BH}}}-1}}\Big]\simeq
{{{4\over9}(A\omega^2/4\pi)^4}\over{e^{\hbar\omega/T_{\text{BH}}}-1}}\ll1$
in the low frequency regime $M\omega\ll1$ [see Eq. (\ref{Eq9})]
which characterizes the neutral sector of the Hawking black-hole
emission spectra in the low-temperature (near-extremal) regime
(\ref{Eq8}).
\end{thebibliography}
\end{document}